# Multiscale characterization of interfacial region in flexible rubber composites: initial structure and evolution upon thermal treatment

S.K. Enganati[a,b], F. Addiego[a], J.P.C. Fernandes[a], Y. Koutsawa[a], B. Zielinski[c], D. Ruch[a], G. Mertz[a,*]

[a]Luxembourg Institute of Science and Technology, L-4940 Hautcharage, Luxembourg.
[b]University of Luxembourg, 2 avenue de l'Université, L-4365 Esch-sur-Alzette, Luxembourg.
[c]Goodyear Innovation Center Luxembourg, L-7750 Colmar-Berg, Luxembourg.

[*]Corresponding author. E-mail address: gregory.mertz@list.lu (G. Mertz), Postal address: Luxembourg Institute of Science and Technology, L-4940 Hautcharage, Luxembourg

**Abstract**

To investigate the structural changes upon thermal treatment of resorcinol–formaldehyde–latex (RFL) interfacial layer in rubber-cord flexible composite, a multiscale approach has been employed. High-resolution AFM mapping showed a significant increase of the modulus of the RF phase from 1.2 GPa to 2.3 GPa and latex phase (L) from 0.3 GPa to 0.8 GPa after a thermal exposure at 100°C for 10 days. The increase of the RF and L phases elastic modulus was correlated with the increase of the oxygen content in RFL layer based on the measurements by SEM-EDX. Besides by combining finite element simulations and AFM modulus profiling, the presence of an interphase region (over 280 nm) between the RFL and the rubber regions was identified and was not deteriorated by the thermal treatment. Peel adhesion testing revealed that the increase of RF and L phases rigidity after thermal treatment was detrimental for the interfacial adhesion of the rubber composite.

## 1. Introduction

Polymeric cord-reinforced rubber composites are extensively used as light and flexible materials in various applications. A cord is represented as a bunch of fibers twisted together to maintain the dimensional stability and improve the mechanical properties and durability of materials used in various applications such as tires, belts, and hoses. In tires, the cords are generally not compatible with the rubber matrix due to lack of adhesion, hence special coatings for the cords are developed to improve the cord-matrix adhesion. For this purpose, resorcinol–formaldehyde–latex (RFL) dip is widely employed on the tire cords, due to its low viscosity and good wetting properties [1]. RFL dip is applied to the cords by a dipping process, during which it is transformed into an insoluble system [2]. The latex gives the required flexibility and adhesion, while the RF resin provides a desirable heat and fatigue resistance by forming a three-dimensional molecular network [3-5]. The RFL adhesive ensures a chemical bonding between the cord and the rubber matrix, forming an interfacial layer [6]. During the composite curing, the increase of the temperature allows the vulcanization process to occur and the rubber matrix is cross-linked in the presence of sulfur [7]. Moreover, it is well accepted now that sulfur diffuses from the rubber matrix to the RFL dip and forms sulfur cross-links in the latex part enhancing RFL-rubber adhesion [8].

Indeed, the composite interfacial adhesion has been widely investigated by peeling, H-testing, or T-testing methods, which are generally used for cord-rubber composites adhesion testing [9-14]. Jamshidi *et al.*, [14] has shown that 150 °C is the suitable vulcanization temperature for the nylon cord-rubber systems by conducting H-adhesion testing at various conditions. Moreover, the composite chemical composition has been studied using a scanning electron microscope equipped with energy-dispersive x-ray spectroscopy (SEM-EDX) showing that the RFL dip has a high affinity towards the accelerator molecules and a moderate affinity towards polymeric or cyclic sulfur molecules in the RFL dip-rubber model system [15, 16]. To go deeper into the understanding of the mechanism occurring at the RFL interface, micro-mechanical properties have been investigated by nano-indentation. These studies highlighted an increase of average values when approaching RFL–rubber interface which was attributed to a local rubber curatives enrichment [17, 18].

Accordingly, the existence of an interphase region between the RFL layer and the rubber matrix with graded elastic properties has been considered but has never been proved by an in-depth study. Understanding and controlling such an interphase between the cord and rubber matrix is a crucial criterion driving the fatigue life of the composites [19]. Due to the recent developments of the AFM technique (the extraction of nano-mechanical data from the force-distance curves), the properties at the nanoscale resolution can be explored with an emphasis on the RFL dip interfacial layer and possibly on the potential presence of interphase between the RFL layer and the rubber.

Besides, during usage, a tire is subjected to continuous heat along with fatigue loading, which also has an impact on its performance and durability. The increase in the temperature of H-pull testings in many cord-rubber systems showed a decrease in adhesion [14]. However, to our knowledge, no analysis of the interfacial layer at the nanoscale was carried out on flexible rubber composites to explore this thermal deterioration effect. Nevertheless, some pioneering works using nanoindentation were performed by Valantin *et al.*, [20] to follow the impact of fatigue loading on the interfacial layer elastic modulus. The degradation of the mechanical property during fatigue loading was linked to a strong increase of elastic modulus of both the RFL region and the closest rubber region. This hardening was attributed to chemical diffusion, such as rubber-curing agents (metallic ions acting as catalysis highlighted by time-of-flight-secondary ion mass spectrometry analysis ToF-SIMS), but also thermo-mechanical aging caused by high local stresses [20, 21]. To our best knowledge, the mechanism occurring during thermal treatment of cord-rubber composites and involving the hardening of the RFL has not been clearly identified.

Hence, in this paper, we propose to further explore the structure of the RFL interfacial layer in the case of a simplified flexible rubber composite by a multiscale approach. This composite system is composed of polyamide 6,6 monofilaments embedded in a rubber matrix. The most important experimental part will rely on the use of AFM equipped with a nano-mechanical mode to resolve the elastic modulus of the RFL layer at the nanoscale. AFM results will be discussed based on simulations of the AFM indentation process by finite element (FE) analysis to tentatively prove the existence of an interphase region between the rubber and the RFL layer with graded elastic modulus. A complementary analysis will be performed by using peel adhesion testing and SEM-EDX to fully describe the structural properties of the RFL layer.

This study has been done before and after thermal treatment procedures to investigate the potential structural changes of the RFL layer.

## 2. Experimental
### 2.1. Materials

The composition of the rubber compound used as a composite matrix is shown in Table 1. The compound is composed of a blend of natural rubber (NR) grade TSR10 and synthetic cis-polybutadiene rubber (BR) filled with ASTM N234 carbon black (CB) and various other ingredients. The compound was cured by a sulfur system with benzothiazyl disulfide (MBTS) as an accelerator and zinc oxide (ZnO) as an activator. Polyamide 6,6 (diameter 0.35 mm and 950/1 dtex) monofilaments were used as a simplified reinforcing agent when compared to a cord. The RFL dipping systems were prepared in the Goodyear laboratory (Goodyear Innovation Center Luxembourg, Colmer Berg, Luxembourg). The typical RFL formulations to prepare a dipping subcoat and topcoat are shown in Table 2.

Table 1. Rubber compound composition in parts per hundred of rubber (phr).

| Component | Content (phr) |
|---|---|
| Natural rubber (NR) | 70 |
| Polybutadiene rubber (BR) | 30 |
| Carbon black (CB) | 55 |
| Paraffin wax | 1 |
| Microcrystalline wax | 1 |
| Aromatic oil (TDAE) | 7 |
| Polymerized dihydrotrimethyquinoline | 1 |
| Dimethylbutylphenyl-*p*-phenylenediamine | 2.5 |
| Stearic acid | 2 |
| Zinc oxide (ZnO) | 4 |
| Sulfur | 2.5 |
| Benzothiazyl Disulfide (MBTS) | 0.8 |

Table 2. Typical two-stage RFL adhesive formulation in wt%.

| Component | Content in subcoat (wt%) | Content in topcoat (wt%) |
|---|---|---|
| Resorcinol Formaldehyde resin (RF) | 26.96 | 13.88 |
| Isocyanate resin | 6.45 | - |
| Vinyl Pyridine latex | 59.4 | 41.3 |
| Styrene Butadiene latex | - | 40.77 |
| Carbon black (CB) | 6.27 | 2.88 |
| Catalyst (NaOH) | 0.92 | 1.16 |

## 2.2. Processing, curing, and thermal treatment

The composite system studied in this project is made of an NR/BR blend (along with other ingredients including sulfur, CB, ZnO, etc.) reinforced with polyamide 6,6 monofilaments. Firstly, the polyamide 6,6 monofilaments are coated in a two-step dipping process, initially with a subcoat that contains an isocyanate component to create active adhesion and then with a topcoat to have complete adhesion [22]. The curing of the dip has been performed at 250°C for 120 sec. These dipped monofilaments are placed in parallel between two rubber layers forming a composite strip. At last, the whole structure is bladder cured at 150°C for 20 min under uniform pressure of 100 psi. The polyamide 6,6 monofilament has a diameter of 0.35 mm and each rubber sheet is 1.2 mm thick, leading to 2.75 mm thick composites as shown in Figure 1. The thermal treatment is done at 100°C in an oven with normal atmosphere conditions, where the cured composite samples are kept for 5 and 10 days.

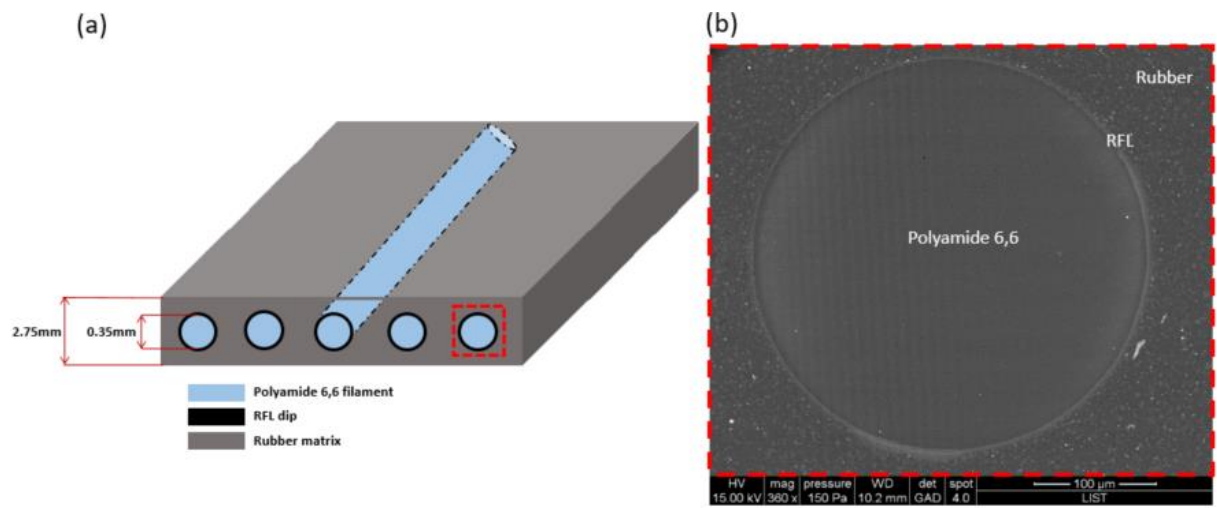

Figure 1. (a) Schematic representation of a composite sheet and (b) SEM image showing a cross-sectional view of a polyamide 6,6 monofilament embedded in the rubber compound.

## 2.3. Sample preparation

Cryo-ultramicrotomy is an advanced sample preparation method to produce a very smooth surfaces for nanoscale resolution analysis. Accordingly, a cryo ultramicrotome Leica EM FC6 was used in our study to prepare our composite, based on a surface trimming step and polishing step. For the surface trimming, a cryo trim 45° knife was used at -120°C inside the cryo chamber to make a truncated pyramid around a single monofilament. Then, for the surface polishing, a cryo-diamond knife 35° was used at a speed of 3 mm/s, slicing off 200 nm thick sections to reach a relatively flat surface on the sample surface. The remaining surfaced block is used for subsequent analysis. Note that a special sample holder was used, enabling to fit with the stage of both the cryo-ultramicrotome and AFM, without removing the sample.

### 2.4. Adhesion peel testing

To measure the adhesive strength, the cords were placed between layers of compound to form a strip. The samples are subsequently cured at $150^0$C for 20 minutes. Then, the strips were partially separated. One rubber ply was clamped in the top grip of an universal testing machine while the other ply was clamped in the bottom grip. Subsequently, 90° peel testing between two the composite plies was performed at a constant extension rate of 50.8 mm/min (testing speed). The peel force required to separate the strip has been recorded. The obtained adhesion strength reflects the interfacial adhesion between the rubber matrix and the monofilaments. Note that this adhesion peel testing preformed here was based on ASTM D4393 Strap Peel Adhesion Test.

### 2.5. Scanning Electron Microscopy with Energy-dispersive X-ray spectroscopy (SEM-EDX)

A SEM model Quanta FEG 200 from FEI coupled with an EDX system GENESIS XM 4i was used to perform SEM-EDX experiments. It was used to focus on the morphology and elemental composition of the RFL interfacial layer. It was operated under low vacuum water vapor inside the chamber, and hence, the composite sample surface can be analyzed without using any conductive coating. EDX analysis was performed across the RFL dip region with an accelerating voltage of 12 kV. The distance between every two-scan points was around 150 nm and with a dwell time of 500 ms.

### 2.6. Atomic Force Microscopy (AFM)

AFM testing was carried out on the RFL interfacial layer with a MFP-3D Infinity AFM from Asylum Research, in fast force mapping (FFM) mode. The tips used were made of silicon with a nominal spring constant of 2 N/m and a nominal radius of 10 nm (model AC240TS from Olympus). All the measurements were made under ambient conditions (room temperature of about 22°C and a relative humidity of about 50%) and a standard cantilever holder for operation in air was used. While scanning the sample surface, the tip oscillated in the Z-direction (travel distance of 300 nm) with a frequency of 300 Hz (300 force-distance curves per second). The images were recorded with 256×256 pixels, giving a lateral resolution of 20 nm for 5×5 µm$^2$ image with a force-distance curve recorded for each pixel. The force setpoint was fixed to around 15-20 nN during the imaging, which implies different indentation depths in each of the phases due to their different mechanical properties.

For the calibration of the tips, first the deflection sensitivity and the spring constant of the cantilevers were determined using the GetReal Automated Probe Calibration feature. The calibrated spring constant values of the different tips used were about 2.1 ± 0.3 N/m. Then a relative calibration was done to determine the tip radius, using a polystyrene/low density polyethylene (PS/LDPE) as a standard sample with known Young's moduli (2.7 GPa for the PS phase and 100 MPa for the LDPE phase). In the rest of the document, Young's modulus or elastic modulus refer to the same information.

In this procedure, the tip radius was adjusted to obtain the Young's modulus value of 2.7 GPa for the PS phase matching with the indentation depth applied in the polyamide 6,6 and RFL phases of the composite. The indentation in LDPE phase was matched to the one applied in the rubber phase to obtain the effective radius of contact. These calibrated tip radii, varying from 25 nm at the soft regions to 5 nm in the stiff ones are used to determine the modulus of the different regions of interest (polyamide 6,6, RFL and rubber).

AFM images and force-distance curves were analyzed using the software Igor (version 6.3). Different nanomechanical models were considered to fit the experimental force-distance curves (retraction curves) and the Derjaguin-Muller-Toropov (DMT) model was the one providing better fits for the three regions of interest. For the model fitting, the contact point is defined by an algorithm from the Igor software. The Raw (X) offset from the force distance data, uses the function GetContactSlope to find the rough estimate of the surface contact point. It performs this procedure by fitting the slope of the deflection versus Zsenser in small chunks (~7 points) until the slope is higher than 0. It then fits the deflection from 20% of force to the contact point and extrapolates that line until it intersects the zero force.

## 2.7. Finite element (FE) analysis

The AFM indentation procedure was simulated by finite element (FE) analysis using the Abaqus commercial FE software with a methodology inspired from previous papers [23, 24]. To represent the AFM tip, a conical indenter (half-angle of 70.3°) was considered with an analytical rigid surface. The tested structure consisted of a polyamide 6,6 monofilament (0.35 mm of diameter) coated by 2 µm of the RFL phase and embedded in 2 mm of a Rubber phase. After many numerical experiments, we came up with a simplified FE structure of the monofilament-RFL-rubber composite that corresponded to a simple parallelepiped structure (length of 6 µm, width of 0.5 µm, and height of 0.25 µm) with the three phases having a length of 2 µm (Figure 2 (a)). The Young's modulus of the polyamide 6,6 phase, RFL dip phase, and the rubber phase was set to 3 GPa, 2 GPa, and 0.05 GPa, respectively. Concerning their Poisson's ratio, they were set to 0.33, 0.33, and 0.5, respectively. A simple elastic behavior was chosen for the polyamide 6,6 and the RFL phases, while an equivalent Neo-Hooke hyper-elastic behavior was used for the rubber phase (with the constants C10 = 8.33 MPa and D1 = 0 MPa$^{-1}$). The FE analysis of the indentation was conducted with the static general step in Abaqus with the geometric nonlinearity activated (NLGEOM = ON). An automatic size incrementation was adopted with a maximum number of increments of 100, an initial increment size of 0.01, a minimum increment size of 1×10$^{-5}$, and a maximum increment size of 0.08. Regarding the solver, the default options were selected. The contact interaction consisted of surface-to-surface contact interaction with small sliding between the structure and the conical indented. The contact tangential behavior relies on a frictionless formulation, while "hard" contact and penalty formulations were used for normal behavior. A pinned boundary condition was applied to the bottom surface of the structure, and for the loading, a displacement of 5 nm was applied to a reference point of the analytical rigid conical indenter. The structure was meshed using the 8-node linear brick (C3D8RH) element (Figure 2 (b)) with reduced integration and hybrid formulation. The Young's modulus of the structure at each indentation point was determined from the load-displacement curve, and in particular from the unloading stage (Figure 2 (c)). The slope of the upper part of the unloading stage $\partial F/\partial h$ corresponds to the contact stiffness S and can be correlated to the reduced elastic modulus $E_r$ (the combined elastic modulus of the tip and the sample) using the model of Olivier and Pharr [25].

$$S = \frac{\partial F}{\partial h} = \frac{2}{\sqrt{\pi}} \cdot E_r \cdot \sqrt{A_c} \qquad (1)$$

where $A_c$ is the contact area. The reduced elastic modulus is expressed as a function of the sample Young's modulus E and Poisson's ratio ν, as well as the indenter elastic modulus $E_i$ and Poisson's ratio $ν_i$:

$$\frac{1}{E_r} = \frac{1-ν_i^2}{E_i} + \frac{1-ν^2}{E} \qquad (2)$$

The contact area is calculated considering a Berkovich indenter:

$$A_c = 24.56 \cdot h_c^2 \qquad (3)$$

where $h_c$ is the contact depth, depending on the indentation depth $h_{max}$, the applied load $F_{max}$, and on a constant ξ depending on the indenter geometry (taken equal to 0.75 for a Berkovich indenter):

$$h_c = h_{max} - ξ \cdot \frac{F_{max}}{S} \qquad (4)$$

This method has been used to determine the elastic modulus of the indented structure by post-processing the output data of the FE simulation using the load versus displacement curves. More importantly, FE simulations were conducted for estimating the mechanically affected zone (MAZ) [23] for a better understanding of the material volume involved during the AFM indentation. This volume is expected to correspond roughly to the penetration depth multiplied by the MAZ. Based on the displacement isovalues resulting from the indentation (Figure 2d), the most important displacement, comprised between 10% and 100%, are localized in the central part of the simulated indentation footprint, while the peripheral section can be neglected. The diameter corresponding to this most representative affected material is named as MAZ.

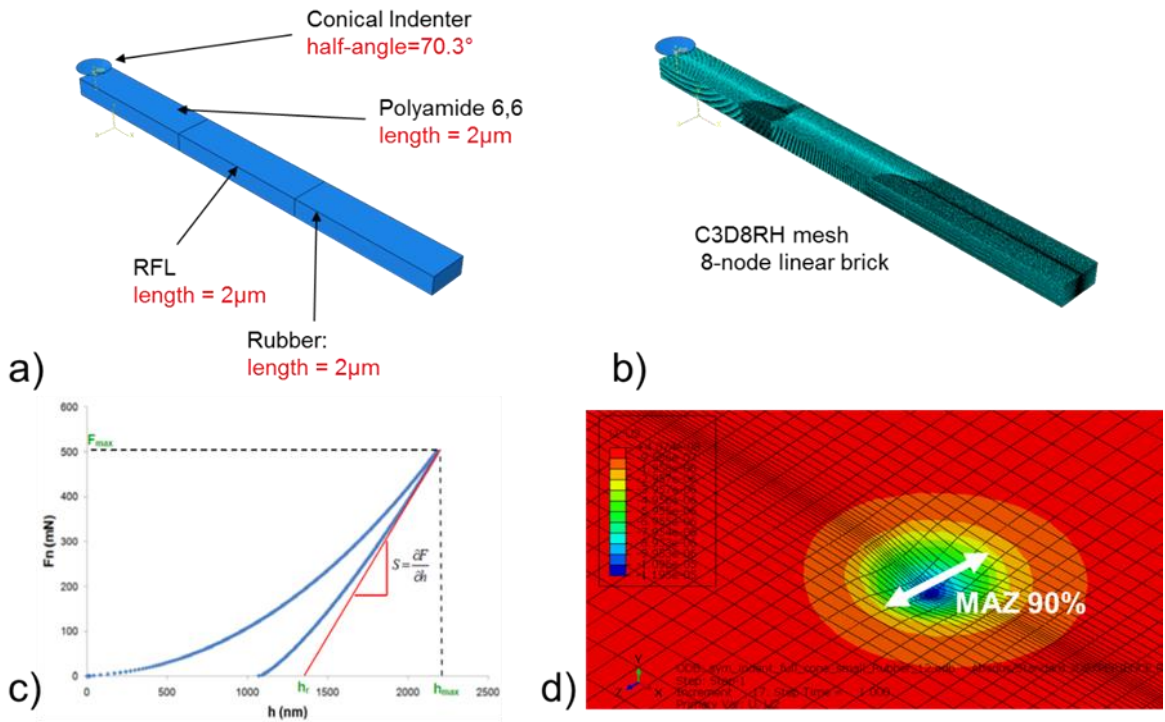

Figure 2. (a) 3 phase model used to simulate the composite (overall length 6 µm, a width of 0.5 µm and height of 0.25 µm), (b) Meshing of the model using C3D8RH mesh in Abaqus, (c) determination of the elastic modulus from the slope of the unloading stage of the force-displacement curve [26], and (d) vertical displacement isovalues in the rubber phase (indentation of 12 nm) enabling to determine the mechanically-activated zone (MAZ) including between 10% and 100% of the displacement [23].

## 3. Results and Discussion
### 3.1. Bulk adhesion properties analysis

The prepared composites sheets are subjected to thermal treatment (at 100°C in the oven) for 5 days and 10 days to determine its effect on the composite's macroscopic properties. The peeling testing was carried out before and after thermal treatments and the obtained adhesion strengths of the samples are tabulated in

Table 3. A decrease in the adhesion strength from 294 N before treatment to 100 N after 5 days of thermal treatment has been observed. The adhesion strength further decreases for the samples thermally-treated for 10 days reaching a value of 36 N. It is likely that this drop in adhesion strength highlights a structural modification which can result in the loss of adhesion between polyamide 6,6 monofilaments and rubber matrix. As discussed in the literature, the loss of macroscopic properties during thermal aging can be explained by a change occurring at the cord/rubber interface and is often associated with the hardening of the RFL layer due to the migration of additives. This paper aims to better understand the macroscale adhesion behavior in the studied composite system and to deeply investigate micro/nanoscale structural changes at RFL dip. For this purpose, further analysis is carried out in the following sections.

Table 3. Peel adhesion testing results of the rubber composites as a function of thermal treatment duration.

|  | Untreated | Treated 5 days | Treated 10 days |
|---|---|---|---|
| Adhesion strength (N) | 294±12 | 100±2 | 36±4 |

### 3.2. Morphological and elemental chemical analysis by SEM-EDX

The SEM imaging of the composite cross-section taken using backscattered electrons (BSE) detection mode is represented in Figure 3(a). The presence of three domains, i.e., rubber, RFL, and polyamide 6,6, can be clearly distinguished in the recorded images due to their difference in terms of topography and chemical composition. The intensity of the backscattered electron is proportional to the mean atomic number of the sample, thus the image contains information on variations in sample composition that makes all the three regions differentiable. The presence of bright spots in Figure 3 (a) can be attributed to the presence of ZnO particles in the rubber matrix. The harder polyamide monofilament presents lines on the surface, most probably due to the cutting process. In between the two regions, the RFL layer is noted with a thickness of about 2 µm. Besides, a rather rough interface between the monofilament and the RFL is observed, while a smooth transition with the rubber is present.

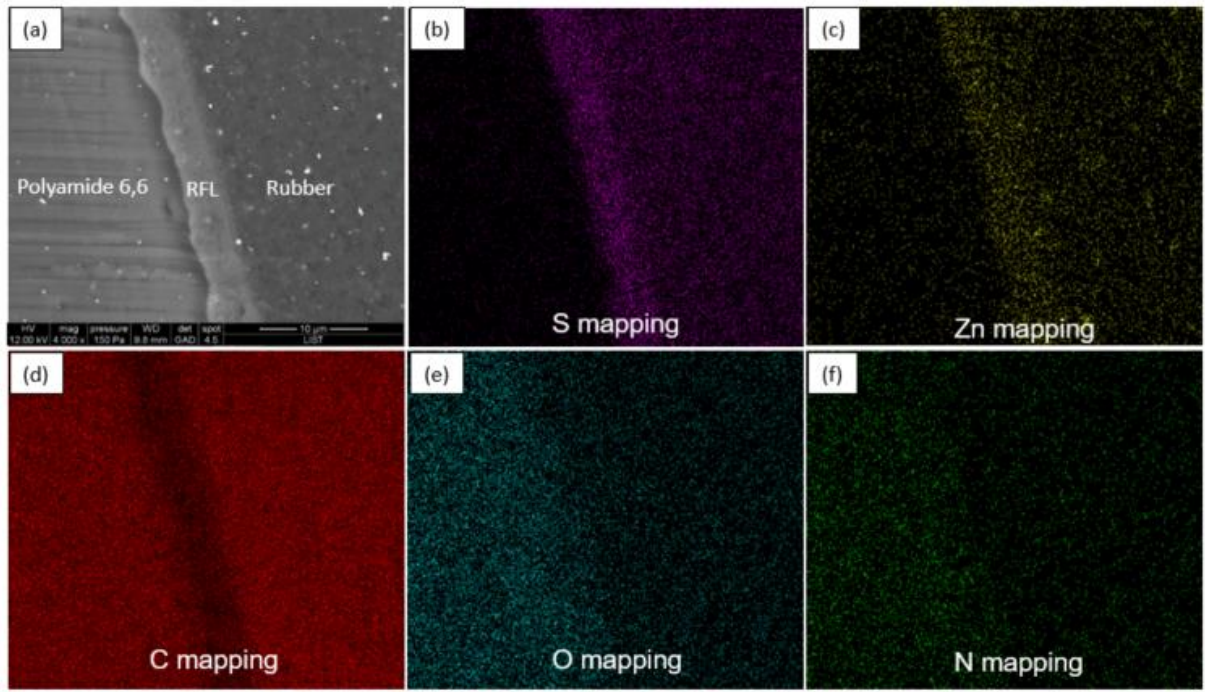

Figure 3. (a) SEM imaging performed on the composite cross-section in backscattered electron detection mode (BSE), and corresponding EDX analyses showing elemental mappings of (b) sulfur, (c) zinc, (d) carbon, (e) oxygen and (f) nitrogen.

The elemental mappings over the selected area are represented from Figure 3 (b) to (f). The higher intensity of the color indicates a higher presence of a given element in the studied region. In the polyamide region, carbon, oxygen, and nitrogen are majorly detected with almost no trace of sulfur and zinc elements. This is consistent with polyamide 6,6 composition. In the RFL dip region, C and O amount detected is less when compared to the polyamide 6,6 region which shows differences in their compositions. From Figure 3 (b) and (c), higher concentrations of sulfur and zinc are observed in the RFL region when compared to the rubber region, which qualitatively confirms the presence of various elements after the curing process.

SEM-EDX scan line measurements have been extracted before and after thermal treatment of the composite samples. Attention is paid to sulfur and oxygen elements, as expected to play an important role during the thermal exposure. As a reference, the carbon element count is also recorded ($C_{count}$). The length of the line is approximately 30 μm and the distance between every two-scan points was around 150 nm, with a dwell time of 500 ms. The various scan lines measured across the dip from rubber to polyamide on untreated, 5 days treated and 10 days treated samples are shown with different colors in Figure 4. The relative sulfur count ($S_{count}/C_{count}$) and relative oxygen count ($O_{count}/C_{count}$) along the scan lines are represented in Figure 4 (a) and Figure 4 (b) respectively. Each elemental count is divided by carbon count to avoid the discrepancy due to the unevenness of the sample surface, as carbon count should be consistent in the respective phases. Indeed, the sulfur count at the RFL dip region cannot be considered explicitly, as it is a relative value and influenced by the presence of other elements. Especially during the curing and the thermal treatment, the presence of oxygen and zinc can affect the absolute values of sulfur count.

Firstly, considering the untreated sample in Figure 4 (a), it can be seen that there is no sulfur present on the polyamide surface. However, a significant sulfur content is found in the RFL dip region, i.e. up to three times higher than in the rubber phase. It is worth noting that there is no sulfur present in the RFL dip solution before the curing process, therefore it is assumed that the sulfur migration observed here is mainly occurring during the curing process. It has been previously observed using EDX maps that sulfur and zinc migrate from the rubber to the RFL layer during the curing process [27]. This migration is probably due to the higher affinity of the unsaturation of the latex in the RFL phase compared to that of the rubber matrix. Importantly, the transition between the rubber region and the RFL dip region is progressive, indicating a potential intermediate phase over few microns (here 2.5 µm), just higher than the typical EDX resolution of 1 µm. This intermediate phase may be an interphase, but further investigations are needed to prove its existence. To this end, AFM and FE have been used in the next sections.

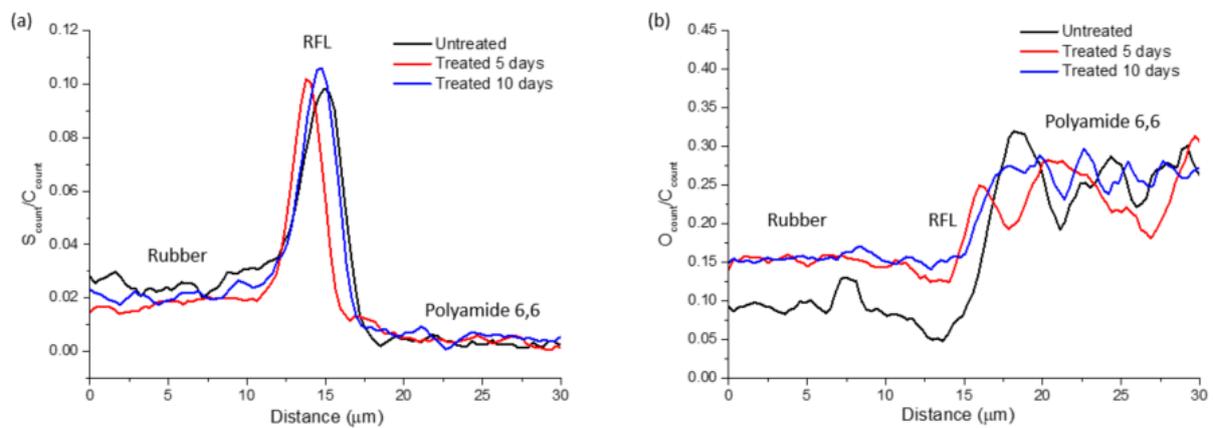

Figure 4. (a) $S_{count}$ / $C_{count}$ (b) $O_{count}$ / $C_{count}$ along the line scan through RFL dip in the case of reference (untreated) sample, 5 days and 10 days thermally treated sample.

After thermal treatment a slight increase in the relative sulfur content at the dip region from untreated to 10 days treated sample appears and could be assigned to the migration of sulfur from the rubber to the RFL phase during the thermal treatment. However, to confirm the sulfur migration from the rubber to the RFL dip region during the thermal treatment a decrease in relative sulfur content in the rubber region from untreated to 10 days treated sample should be observed. Unfortunately, due to the inhomogeneity of the materials, the sensitivity of the SEM-EDX technique and the presence of other elements such as oxygen during thermal treatment provide some uncertainties to conclude on a potential migration of sulfur element. On the other hand, it is observed on Figure 4 (b) that the $O_{count}/C_{count}$ ratio increases with increasing the thermal treatment time, from 0.05 in the case of reference (untreated sample) to 0.12 after 5 days and up to 0.15 after 10 days. This finding indicates the presence of more oxygen in the dip and rubber matrix after thermal treatment, increasing the content of oxygen of both materials. Such an increase in oxygen content can lead to an increased cross-link density in the RFL dip region, which tends to become more rigid and thus prone to premature damage. This could be a possible reason of the decreased adhesion strength after thermal exposure (Table 3). To bridge such an understanding between the elemental chemical behavior and macroscopic mechanical behavior, nanoscale mechanical analysis is considered and performed as shown in the following section.

## 3.3. Morphological and nano-mechanical analysis by AFM

AFM analysis was performed on the reference composite system to obtain quantitative mechanical properties at the nanoscale such as the Young's or elastic modulus. The topography view of a 5×5μm$^2$ area is shown in Figure 5 (a). As for the SEM images, the three components of the composite can be distinguished even using the topography images. The polyamide region with a Ra of 10 nm appears smoother than the other components. The rubber region appears much rougher (Ra = 33 nm), mainly due to the presence of various compound ingredients (ZnO particles and carbon black, for instance) embedded in the NR/BR blend. The RFL dip region, about 2 μm thick, appears with an intermediary roughness, with Ra = 21 nm. When the AFM tip approaches the sample, the rubber, RFL dip, and polyamide domains interact differently with the tip due to the difference in their intrinsic properties. Therefore, mechanical properties can be simultaneously acquired with topography. The modulus mapping overlapped on topography is shown in Figure 5 (b). The contrast in mechanical properties clearly distinguishes the components, with the rubber region appearing with the lower modulus level (blue color in the image), the RFL dip region with an intermediary modulus (blue/red), and the polyamide with the highest modulus level (yellow color).

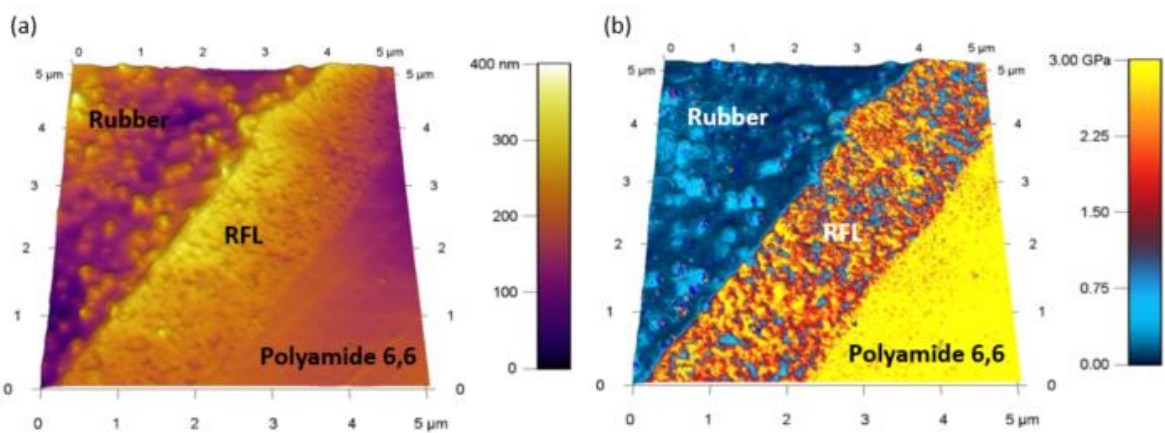

Figure 5. 5×5μm$^2$ AFM imaging focused on the three regions (rubber matrix, dip, and monofilament) of the untreated composite sample showing topography mapping (a) and modulus mapping (b).

The modulus is directly extracted from the force-distance curves provided at each interaction of the AFM tip with the surface of the sample. A typical force-distance curve is shown in Figure 6, highlighting a full cycle characterized by an approach curve and a retract curve. The cantilever starts from a free level position in 'A' and moves towards the surface. At the position 'B' the cantilever is attracted to the surface without a real load on it due to van der Waals forces. While the cantilever is moving further into the sample, it starts to bend with an increasing loading force. When the cantilever reaches the maximum predefined loading force, the approach curve ends and the retract curve starts. The cantilever moves in a reversed direction to move away from the sample. Due to the adhesion, the tip may not leave the surface at the same position 'B'. The cantilever remains in contact with the surface until the maximum adhesion force is overcome and then the tip is detached from the sample. The yellow shaded region in Figure 6 corresponds to the work of adhesion. Depending on the material being in contact with the tip the force-distance behavior differs and is related to its intrinsic properties [28].

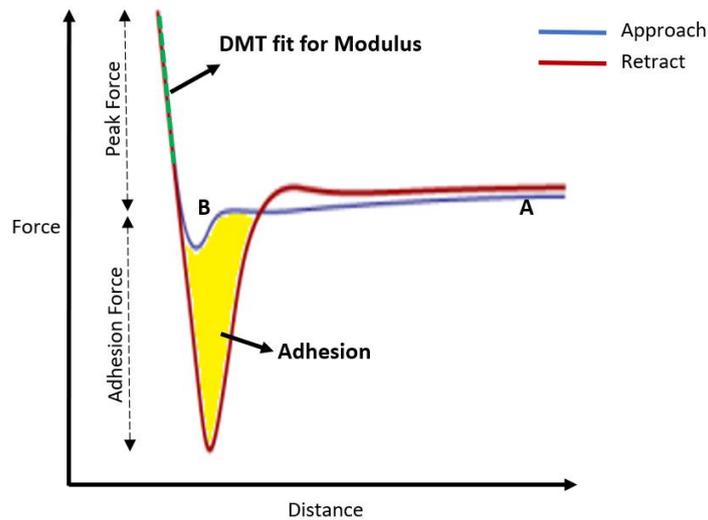

Figure 6. Schematic illustration of a force vs. distance curve during AFM mapping. The blue line indicates the process of the tip approaching the sample, while the red line indicates the process of the tip retracting from the sample.

As three distinct regions composed the studied composites, the difference in these material's behavior can be observed from their force-indentation curves. The force-indentation curves at two different locations in the polyamide region are presented in Figure 7. The dotted lines correspond to the approach curves and the solid lines correspond to retract curves. It can be observed that the trajectory of force-indentation curves of the points 1 and 2 are very similar indicating an homogeneous behavior. The average adhesion force is around 15 nN for the polyamide region.

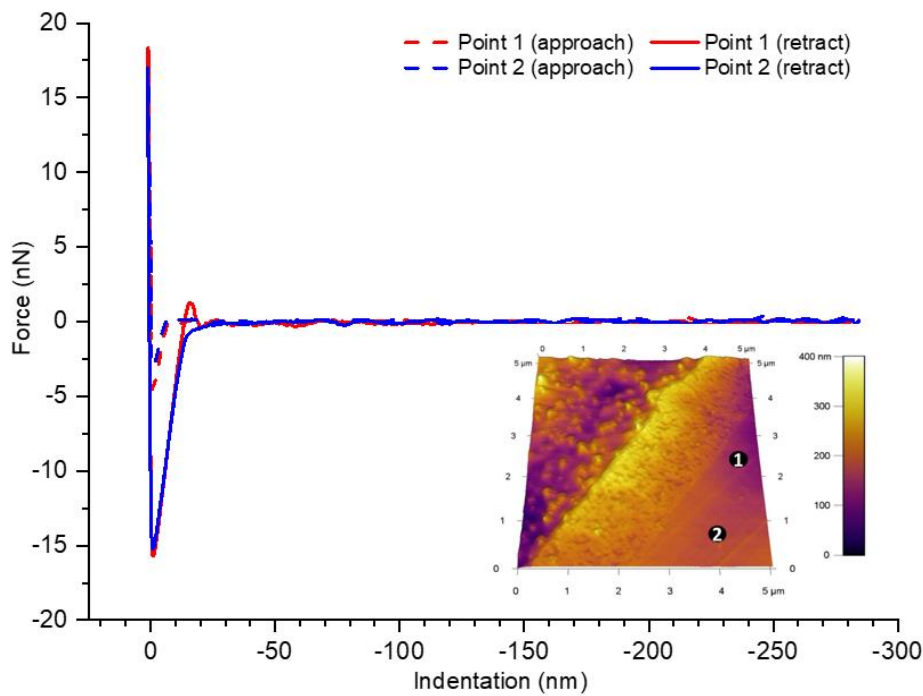

Figure 7. Measured force-indentation curves by AFM in the monofilament region of the untreated composite sample. The dotted lines indicate the process of the tip approaching the sample, while the solid lines indicate the process of the tip retracting from the sample.

The force-indentation curves at four different locations, two in the RFL dip and two in the rubber region are presented in Figure 8. It can be observed that the point 3 and point 4 in the RFL have different behaviors, mainly from the adhesion force and work of adhesion. Point 3 has a relatively lower work of adhesion compared to the point 4, indicating that the point 3 is comparable to polyamide behavior. When approaching the rubber region, the trajectory of force-indentation curves of the points 5 and 6 are very similar indicating the consistency in material behavior. The point 4 in the RFL region can be related to rubber behavior (points 5 and 6). Thus, the force-distance curve at the point 3 could correspond to the RF resin phase and the curve obtained at point 4 is in the latex phase of the RFL dip region. Besides, from such curves (Figure 8) the adhesion forces can be extracted and computed to ~40 nN in the rubber phase. Further, it can be pointed out that the adhesion forces of the latex phase (point 4) and rubber region (points 5 and 6) are similar but their work of adhesion trajectory is completely different highlighting the high sensitivity of the technique. It is also important to note that approach and retract curves from polyamide 6,6 and RFL regions superpose well indicating an elastic contact. However, we could see the difference in slopes in rubber region force curves, due to a possible limitation of the modulus estimation in such regions. Using these force curves, the modulus can be extracted and computed using the Derjaguin-Muller-Toropov (DMT) model. Indeed, the modulus is determined by fitting the green dashed line in Figure 6 to DMT model using the following equation:

$$F = \frac{4}{3} \frac{E}{(1-\nu^2)} \sqrt{R} \delta^{\frac{3}{2}} - 2\pi R \gamma \qquad (5)$$

Where F is the applied force (from the force-indentation curve), E corresponds to Young's modulus (fit parameter), ν is Poisson's ratio (typically 0.3-0.5 for polymer materials), R is the radius of curvature of the tip, δ is indentation depth and γ the work of adhesion [29].

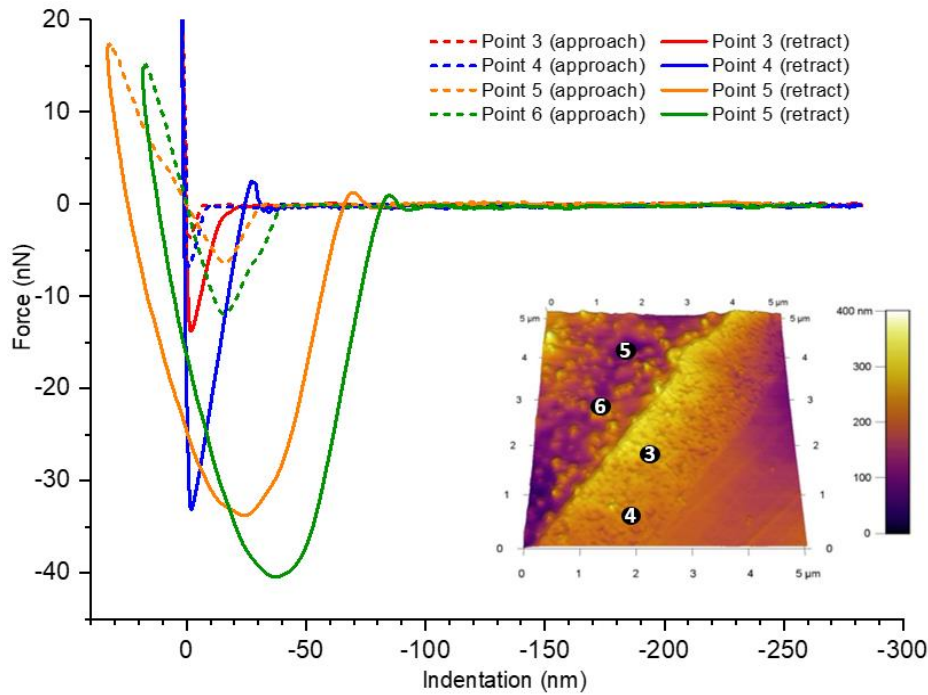

Figure 8. Measured force-indentation curves by AFM in the RFL dip and rubber region of the untreated composite sample. The dotted lines indicate the process of the tip approaching the sample, while the solid lines indicate the process of the tip retracting from the sample.

From the force-indentation curves and after calibration of the tip, quantitative analysis is done giving an average elastic modulus (E) in the case of the polyamide, RFL dip, and rubber regions of 3.0 ± 0.3 GPa, 2.0 ± 0.5 GPa and 50 ± 25 MPa, respectively (Figure 9 (a)). The uncertainty stated here is the average of various images measured on the different locations of the samples with different tips. The Modulus of each region in the AFM images were measured by masking out the image for that region only, and then calculating the average and standard deviation. The reported values consider at least 4 images in each sample for more reliable results. According to the obtained AFM results, the polyamide domain seems to be very smooth with a relatively homogeneous elastic modulus distribution, which is not the case for the dip and rubber domains. The morphology of the rubber domain is composed of different phases, i.e., the immiscible NR and BR phases, but also hard ZnO and CB particles [30]. The low amount of the BR phase in the composition and its similar mechanical properties to the NR phase does not allow us to distinguish these two phases. Such an inhomogeneity in the rubber region is majorly responsible for the large uncertainty observed of the average modulus value. Furthermore, the complex morphology of the RFL dip is revealed. From Figure 9 (a) it can be observed that the dip is composed of a highly cross-linked RF resin, assigned to a yellowish-red color and a latex part, which is known to be softer is represented by a bluish color, similar to the rubber phase.

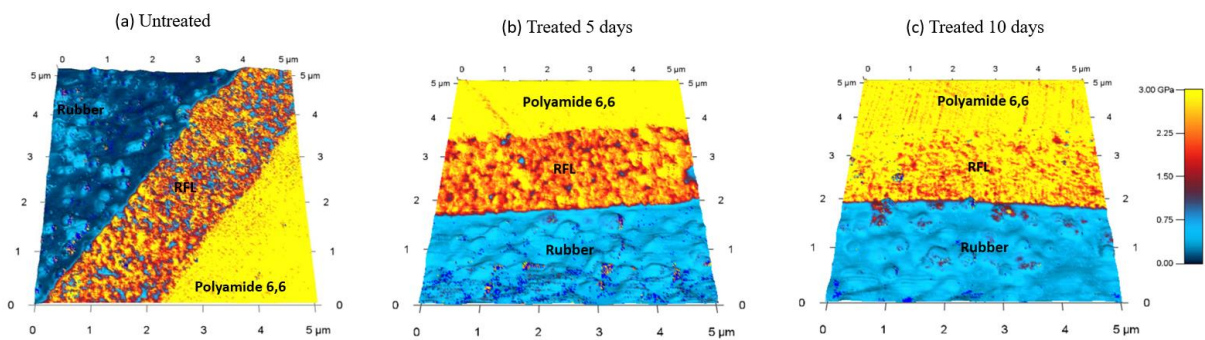

Figure 9. 5×5μm$^2$ AFM imaging focused on the three regions (rubber matrix, RFL dip, and polyamide 6,6) showing modulus mapping of (a) reference (untreated sample) (b) 5 days of a thermally treated sample, and (c) 10 days of thermally treated sample.

Related to thermally treated samples, AFM measurements are performed close to the same areas before and after thermal treatment to avoid artefacts and have a sensible comparison. The elastic modulus mapping overlaid on the topography of thermally treated samples is presented in Figure 9 (b) and (c). It can be observed that the elastic modulus of the RFL dip region is progressively increased reaching almost the elastic modulus of the polyamide region after 10 days of thermal treatment. The same trend is observed in the rubber phase with an increase of the modulus with the thermal treatment. No modification seems to occur in the case of the polyamide phase. Such an increase of elastic modulus in the RFL dip is correlated to an increase of hardness which is in a good agreement with the observations of the literature [21]. This result tends to illustrate that the dip region is becoming more and more hard during the thermal treatment, losing its intermediate elasticity between that of the rubber phase and that of the polyamide phase. According to the literature [17], one reason to explain the inherent RFL dip hardening during the treatment is due to a growth of the three-dimensional resin molecular network. Besides, based on the reported SEM-EDX results, an increase of the presence of oxygen is noted in the dip. It is assumed that during the thermal treatment, a combination of oxidation and possible sulfur cross-links in the dip region could lead to an increase in the overall cross-link density. Therefore, considering both hardening and cross-linking phenomena during thermal treatment may result in a more rigid molecular network of the RFL region, negatively impacting the adhesion strength at the macro-scale (Table 3). It is believed that the increased rigidity of the RFL interfacial layer leads to a poor stress-transfer between the rubber and the polyamide 6,6 monofilament during the peeling testing, resulting in important interfacial damage and hence low adhesion strength. At the state of our knowledge hardening of the RFL layer has been demonstrated but no clear evidence at the nanoscale has been found. Indeed, Valantin *et al.* [21] have combined various techniques to understand the fatigue damage mechanisms of RFL coated textile rubber composites. From the SEM observations, the damage observed was attributed to the propagation of pre-existing fibrillar micro-cracks and the rubber/RFL interface and adhesive debonding between the textile microfilaments and the RFL. Here, there is a lack of information on the nanoscale behavior in or around the RFL region. Hence, to gain a better understanding of the change occurring in the RFL layer during thermal treatment, higher resolution AFM images have been recorded to determine the intrinsic properties of the RF resin and latex parts separately. The dip modulus mapping of 500×500 nm$^2$ area of the samples before and after treating for 5 days and 10 days is presented in Figure 10. The structure of the cured RFL dip consists of a continuous resin phase and dispersed latex particles, which can be observed through AFM images due to the difference in elastic modulus, for instance as shown in Figure 10 (a). The yellow-red color represents the RF resin phase and the bluish color represents the latex phase. During the thermal treatment, the blue region is almost turned into reddish color indicating the increase in elastic modulus in the latex phase Figure 10 (b) and (c).

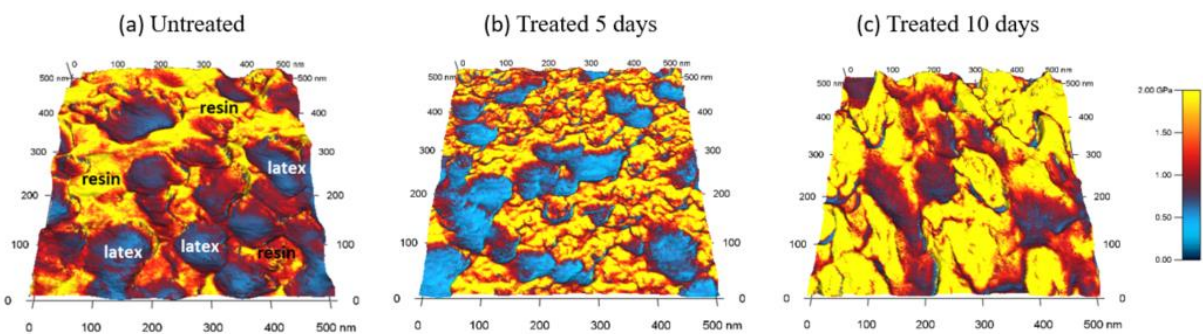

Figure 10. 500×500 nm$^2$ AFM imaging focused on the RFL dip region showing modulus mapping in the dip region of (a) reference (untreated sample) (b) thermally treated sample for 5 days and (c) for 10 days.

The average elastic modulus values of RF resin and latex phases as a function of thermal treatment duration were determined and tabulated in Table 4. It can be observed that the elastic modulus of RF resin phase is gradually increasing from 1.2 GPa up to 2.3 GPa after the thermal treatment which is mostly attributed to resin hardening (self-curing of the resin over time).

Indeed, as hypothesized by Valantin *et al.,* post-reaction of RF by condensation can occur [21]. On the other side, the latex elastic modulus has a slight increase from untreated samples (0.3 ± 0.1 GPa) to the 5 days thermally treated sample (0.4 ± 0.1 GPa). However, it has increased up to two times for the 10 days thermally treated sample (0.7 ± 0.2 GPa). The increase in the latex modulus is probably due to the cross-linking of sulfur migrating from the rubber to the RFL dip region or free sulfur initially present in the dip, or due to thermal oxidation observed for the rubber matrix, as indirectly noted in SEM-EDX measurements (Figure 4). This result clearly shows that the contribution of the increment in the overall RFL dip modulus is not only due to the resin hardening but also to a change in the latex phase as a result of a cross-linking phenomenon. To our best knowledge, this is the first time that the individual contribution of the latex phase and RF resin phase at the nanoscale and its evolution during thermal treatment is reported. Indeed, both phases are subjected and affected during the thermal treatment contributing to the hardening of the layer. During such a thermal exposure, the RFL layer becomes brittle and leads to a loss of adhesion as observed by the peel adhesion test (

Table *3*). The proposed multiscale methodology can pave the way to the improvement of the latex formulations, RF resin ratios, and latex to RF resin ratios to provide a better adhesion and durability.

Table 4. Nano-mechanical properties of the RF resin phase and latex phase as a function of thermal treatment duration.

| In the dip region | Average elastic modulus (GPa) | | |
|---|---|---|---|
| | Untreated | Treated 5 days | Treated 10 days |
| RF resin phase | 1.2±0.1 | 1.6±0.2 | 2.3±0.2 |
| Latex phase | 0.3±0.1 | 0.4±0.1 | 0.8±0.2 |

Additionally, to have more insights into the cord/rubber interface a close look has been done on the modulus profile provided by AFM measurement (see Figure 11(a) and specifically at the rubber-RFL and RFL-polyamide interfaces. A clear sharp interphase occurs between polyamide and RFL, while a graded interphase over a few hundred nanometers is highlighted between the RFL layer and rubber matrix. FE simulation has been attempted to better understand and confirm such an interphase. Besides, beyond the use of FE to investigate the existing interphase, the knowledge provided from such experiments will be beneficial for future improvement of the properties of the RFL layer and its adhesion properties towards fiber and rubber matrix. The use of FE combined with cutting edge analytical techniques could be the key to improve the performance of materials without resorting to numerous and costly experiments.

### 3.4. Nano-mechanical profiling by AFM and FE simulation

The first input data added to the model to predict the modulus profile of the structure have been extracted from the AFM profile itself and set to a constant value for each phase such as 3, 2, and 0.05 GPa for polyamide, RFL layer, and rubber, respectively. The method is described in detail in the experimental part. Briefly, the elastic modulus of the simulated structure is determined based on the post-processing of the output data of the FE simulation. The result of the simulation (curve FE-1 untreated) was superposed to the experimental curve in Figure 11(a) (here untreated referred to data before thermal treatment). A perfect correlation is seen between the experimental and predicted elastic modulus profile of the untreated composite for polyamide phase, in the polyamide-RFL interface region, and the RFL phase. However, the correlation was not optimal at the RFL-rubber interface, especially at the rubber phase side (Figure 11(b)). Indeed, an increase of the elastic modulus over 280 nm from the rubber to the RFL was noted in the experimental data from the AFM profile but not in the FE analysis. By using constant values for the three phases, the FE simulation does not fit with the experimental data. Indeed, the model seems too simple suggesting the presence of a fourth phase that needs to be refined. To improve the simulation, a fourth phase between RFL and rubber was added to the existing structure (Polyamide, RFL, and rubber matrix). The values of this fourth phase have been extracted from the part of the experimental curve (where the modulus decreases from the RFL to the rubber) and applied to the data processing of the FE simulation (FE-2 curve in Figure 11 (a) and (b)). By adding this fourth phase, the simulation led to a good fitting with the AFM experimental data. This finding supports and argues in favor of the existence of an interphase region between the RFL and rubber matrix with a gradient of elastic modulus. This interphase may be related to the diffusion of elements such as zinc and sulfur between the RFL and the rubber during the processing such as highlighted by Wennekes [17] and Valantin et al.,[20] and confirmed by SEM profiling (Figure 4). This interphase possesses intrinsic properties different from the rubber and the RFL, like a mixture between both phases. This last comment opens the door to the hypothesis of a possible intermingling between the RFL and rubber matrix taking place during the curing. Indeed, the high temperature and pressure reached during the composite processing could enable the intermingling between both materials. Unfortunately, such a hypothesis of the intermingling of the RFL part to rubber or vice-versa needs further investigations and will be the subject of future research. Moreover, the same approach has been applied to samples after thermal treatment (Figure 11 (c) and (d)). Experimental and predicted data are correlated for both thermal treatments, the shape and the presence of the interphase are similar in all the cases. So, the interphase is not deteriorated during the thermal exposure. The only difference occurs in the value of the modulus of the RFL layer and rubber which are higher compared to the sample before treatment as explained previously by the hardening mechanisms.

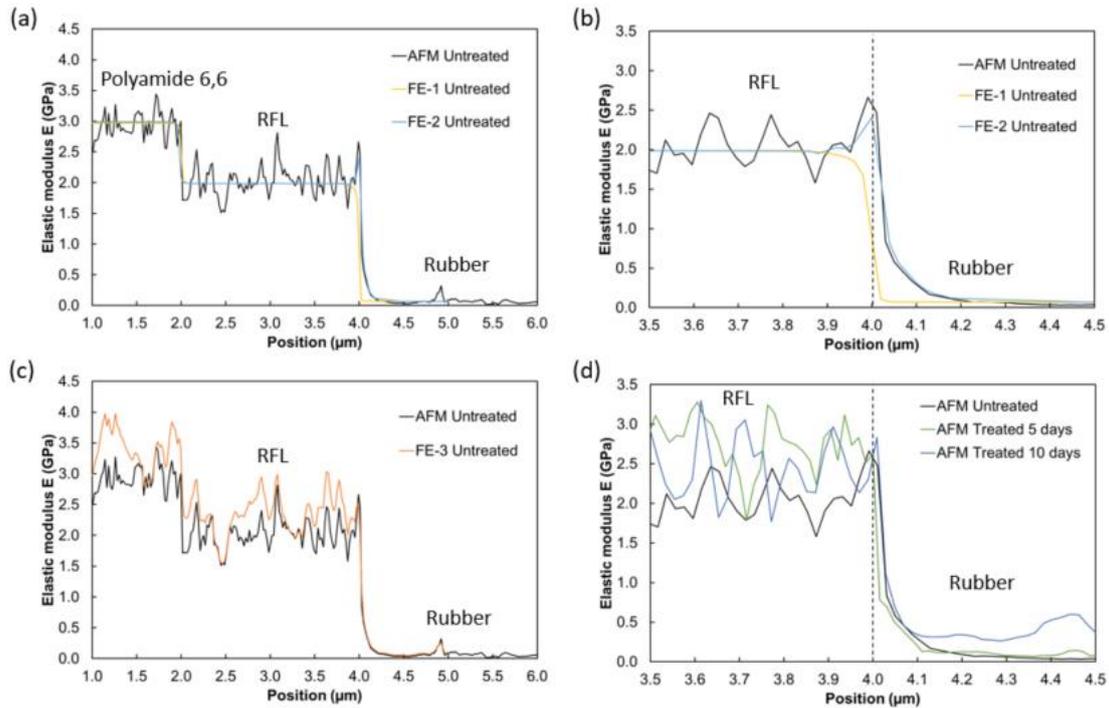

Figure 11. (a) Correlation between the experimental AFM elastic modulus profile and the FE simulation (FE-1 condition with constant elastic modulus in each phase, and FE-2 condition is FE-1 with a gradient of modulus over few hundreds of nanometres at the RFL-rubber interface in the case of the untreated composite, (b) the same graph as (a) with focus on the RFL-rubber interface, (c) correlation between the experimental AFM elastic modulus profile and the predicted one by FE analysis (FE-3 condition with the real AFM profile as input data) in the case of the untreated composite, and (d) effect of thermal treatment on the AFM elastic modulus profile with a focus on the RFL-rubber interface.

To further understand the AFM indentation, and in particular the involved material volume during the indentation, the mechanically-activated zone (MAZ) was determined in the rubber and in the RFL phases varying the applied tip displacement from 4 to 16 nm (Figure 12). Similar MAZ values were obtained for the two regions, and for an applied tip displacement of 5 nm, a MAZ of 50 nm was obtained. This finding implies that the volume involved during the AFM indentation is approximately a cylinder of diameter 50 nm and depth of 5 nm. Therefore, the identified interphase region with a size of 280 nm is realistic considering the resolution of the indentation method.

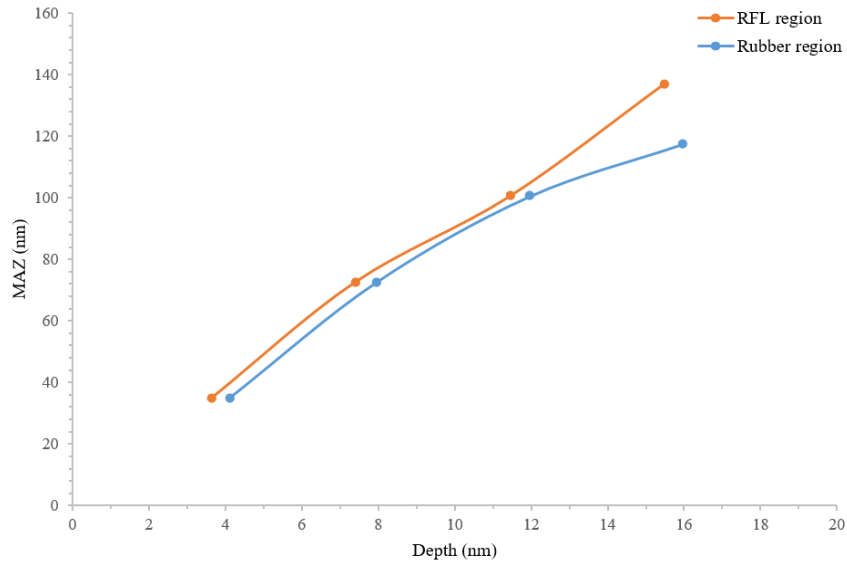

Figure 12. Evolution of the MAZ as a function of the appplied indentation depth from 4 to 16 nm determined by FE analysis simulating AFM indentation.

## 4. Conclusions

In flexible cord-reinforced rubber composites, the adhesion between the cord and the rubber matrix is ensured by a resorcinol-formaldehyde-latex (RFL) interfacial layer. This layer plays an important role in the composite functionality and durability, but its detailed structure has not been resolved at the nanoscale. The current works aim at resolving the RFL layer structure in the case of polyamide 6,6 monofilaments embedded in a rubber matrix through advanced AFM measurements. In particular, the elastic modulus of the RFL interfacial adhesive layer at the nanoscale is determined by AFM tip indentation and discussed based on simulations by finite element (FE). Complementary characterizations of the RFL layer structure are done by peel adhesion testing and SEM-EDX.

AFM profiling revealed the elastic modulus of each component that were 3 GPa for polyamide 6,6 phase, 2 GPa for the RFL layer and finally 0.05 GPa for the rubber phase. Besides, FE simulations provided a mechanically-activated zone (MAZ) of around 50 nm for an applied displacement of 5 nm. This finding demonstrates that the involved volume in the nano-mechanical measurement was roughly a cylinder of diameter 50 nm and height of 5 nm. Using FE simulations of the AFM indentation procedure, we noted the existence of an interphase between the RFL layer and the rubber phase with a gradient of elastic modulus over 280 nm. This interphase was larger than the involved volume, confirming its existence. Besides, thermal exposure did not deteriorate this interphase those size and elastic modulus remain constant. When submitted to thermal treatment at 100°C, the ratio between the oxygen count and the carbon count of the RFL region increased from initially 0.05 to 0.15 after 10 days. At the same time, from the high-resolution AFM imaging, an increase in the elastic modulus from 1.2 GPa to 2.3 GPa in the RF phase and 0.4 GPa to 0.8 GPa in the latex phase was observed after thermal treatment. Both the RF three-dimensional molecular network and latex phase have an increased elastic modulus based on AFM indentation measurements. It was explained that the increase in oxygen content in the RFL region is probably responsible for an increase in the cross-links density of the RF and latex phases. Thermal treatment also induced a resin hardening in the case of the RF molecular network. These two mechanisms may explain the increased elastic modulus of the RFL region, appearing detrimental for the interfacial adhesion as revealed by peeling testing.

However, the exact chemical structure evolution of the RFL layer engendered by the thermal treatment has not been identified yet and will be the scope of future works. It is anticipated that the coupling of AFM with infrared spectroscopy may address this research need. All the findings of this work may contribute to improve the compatibilization technology between the reinforcing agent and the rubber matrix for more durable flexible composites.

**Acknowledgment**

The authors are grateful to Fonds National de la Recherche Luxembourg (FNR) for funding the project (project reference IPBG16/11514551/ TireMat-Tech). The authors are also grateful to Goodyear Innovation Center Luxembourg for their continuous help and support. Sachin Kumar Enganati shows his gratitude to Asmaa El Moul and Jean-Luc-Biagi from LIST for their great technical support.